\documentclass[12pt]{article}
\usepackage{amsmath,latexsym,amssymb,graphicx,axodraw}

\makeatletter

 \@addtoreset{equation}{section}
\makeatother

\renewcommand{\thefootnote}{\roman{footnote}}

\newcommand{\vs}[1]{\vspace*{#1 mm}}
\newcommand{\hs}[1]{\hspace*{#1 mm}}

\begin{document}

\vfill\eject
\begin{titlepage}
\begin{flushright}
OU-HET 463

hep-ph/0312358

April 6, 2004 
\end{flushright}

\vskip1.5cm
\begin{center}
{\Large{\bf
Extension to the Super-Weyl-K{\" a}hler Symmetry in~Anomaly~Mediation
}}
\end{center}

\vskip2cm
\begin{center}
{\large {\bf Koske Nishihara}}\footnote{e-mail:koske@het.phys.sci.osaka-u.ac.jp}
{\large {\bf and Takahiro Kubota}}\footnote{e-mail:kubota@het.phys.sci.osaka-u.ac.jp}

\vskip0.2cm
{\it Graduate School of Science, Osaka University, Toyonaka, 
Osaka 560-0043, Japan}
\end{center}

\vskip2cm
\begin{abstract}
Anomaly mediation mechanism of supersymmetry breaking is known to have an unsatisfactory aspect which is often referred to as the tachyonic slepton problem. A possible way out of this difficulty is investigated by considering the mediation mechanism not only of the super-Weyl but of K{\" a}hler part of anomaly. A new formula for the scalar mass squared is presented by including the effect of the K{\" a}hler anomaly. On the basis of the new formulae, the gauge non-singlet scalar mass squared is shown to be positive provided that the superpotential in the hidden sector satisfies a certain condition. The gauge singlet scalar mass squared may also be positive if we consider the threshold correction.
\\ \\
{\it Key words:} anomaly mediation, super-Weyl-K{\" a}hler symmetry, slepton mass
\\ 
{\it PACS:} 04.65.+e,11.30.Pb,12.60.Jv

\end{abstract}
\end{titlepage}

\setcounter{footnote}{0}
\renewcommand{\thefootnote}{\arabic{footnote}}

\section{Introduction}
It has been believed that the Minimal Supersymmetric Standard Model (MSSM) is one of the promising candidate for the theory beyond the Standard Model. The real world, however, is not given apparent manifestation of the supersymmetry (SUSY), so that this symmetry should be broken down somehow.  Various scenarios for a SUSY breaking mechanism have been studied so far, such as gravity mediation \cite{Nilles:1983ge}, gauge mediation  \cite{Giudice:1998bp} and so forth.

Several years ago, a novel type of SUSY breaking mechanism was proposed based on the super-Weyl anomaly which is often referred to briefly as anomaly mediation mechanism \cite{Giudice:1998xp}\cite{Randall:1998uk}. The most appealing aspect of anomaly mediation is its unique predictability of soft breaking terms (SBT's). For example gaugino masses, scalar masses  and triple coupling of matter multiplets are given respectively by
\begin{eqnarray}
M_\lambda & =& \frac{\beta_g}{g}  \frac{M}{3} ^* \label{AMgauginomass},\\
m^2_i  &=&  -\frac{1}{4}   \left(\frac{\partial\gamma_i}{\partial g}\beta_g,
               +\frac{\partial\gamma_i}{\partial y}\beta_y  \right) \left| \frac{M}{3}\right|^2 \label{AMscalarmass},\\
A^{ijk}  & =&   \frac{1}{2} (\gamma_i+\gamma_j+\gamma_k)y^{ijk} \frac{M}{3} ^* .
\end{eqnarray}
Here indices $i$, $j$ and $k$ label the chiral matter multiplets. These parameters are all determined in terms of the  beta functions of the gauge coupling $\beta _{g}$, the Yukawa coupling $\beta _{y}$ and the anomalous dimensions $\gamma _{i}$ of the $i-$th chiral multiplet  field.  $M$ is the auxiliary field of the gravity multiplet whose vacuum expectation value (v.e.v.) is the source of SUSY breaking.

In spite of its high predictability, the anomaly mediation scenario is not satisfactory in all of its details. Namely, scalar partners of leptons are given  negative mass squared. Various ideas have been proposed to solve this tachyonic slepton mass problem.
They are, however, necessarily to introduce additional assumptions which include: (i) new bulk field interactions in extra dimensions \cite{Randall:1998uk}\cite{Rattazzi:2003rj}, (ii) threshold effects due to new gauge interactions \cite{Pomarol:1999ie}-\cite{Katz:1999uw}, (iii) extra Yukawa couplings \cite{Allanach:2000gu}\cite{Chacko:1999am}, (iv) Fayet-Iliopoulos $D$ terms of U(1) gauge interactions \cite{Jack:2000cd}-\cite{Luty:1999qc} and (v) horizontal gauge symmetry \cite{Anoka:2003kn}.
All the solutions of the tachyonic slepton problem are of course very interesting in their own right. 
In this paper, however, we would like to  stick to the simplicity of the original anomaly mediation as much as possible. We extend the super-Weyl to the super-Weyl-K{\" a}hler anomaly and pursue the possibility to solve the slepton mass problem with the help of the additional K{\" a}hler contribution. In the course of our investigation of the singlet scalar case, we find it necessary to evade the insensitivity of the anomaly mediation to ultraviolet physics along the line of Nelson and Weiner \cite{Nelson:2002sa}.

\section{Super-Weyl-K{\" a}hler Anomaly}
The minimal supergravity coupled with matter and gauge multiplets is invariant under the super-Weyl-K{\" a}hler transformation on the classical level. On the quantum level this symmetry is broken down, so the theory is anomalous \cite{LopesCardoso:1993sq}. The quantum level anomaly associated with this symmetry contains not only the auxiliary field of the gravity multiplet, $M$, but also the K{\" a}hler potential  on the order of $\kappa ^{2}=8\pi G_{N}$.  In the next paragraph we recapitulate the super-Weyl anomaly contribution driven by the v.e.v. of $M$, thereby reproducing diagrammatically the SBT's. Our graphical analysis makes it easier to derive additional SBT's which originate from the K{\" a}hler potential. 
This idea considering these additional contributions was put forward for the case of the gaugino term in Ref.\cite{Bagger:1999rd} and for several SBT's in Ref.\cite{Gaillard:2000fk}.
Gaillard and Nelson \cite{Gaillard:2000fk} in particular made use of the K{\" a}hler U(1) superspace formalism \cite{Binetruy:2000zx} to compute one-loop quantum effect. 
These additional terms could  hopefully change the nature of the tachyonic slepton problem. 
In the following, we resume this idea from phenomenological viewpoints to calculate additional contribution to the scalar mass terms coming from K{\" a}hler part at the two-loop level.

\subsection{Weyl part of Super-Weyl-K{\" a}hler Anomaly}
\begin{figure}[t]
\begin{center}
\begin{picture}(500,80)(20,25)
\SetWidth{1.0}
\Photon(80,87)(80,97){2.5}{1}
\Photon(30,27)(50,37){2.5}{2}
\Photon(110,37)(130,27){2.5}{2}
\Line(50,37)(110,37)
\Line(50,37)(80,87)
\Line(110,37)(80,87)
\Text(75,107)[lb]{\large{$R^*$}}
\Text(135,12)[lb]{\large{$V$}}
\Text(15,12)[lb]{\large{$V$}}
\Text(105,62)[lb]{\large{$Q,Q'$}}
\Text(72,6)[lb]{\large{$\mbox{(a)}$}}
\Photon(250,87)(250,97){2.5}{1}
\Photon(280,37)(300,27){2.5}{2}
\Photon(220,37)(200,27){2.5}{2}
\ArrowLine(220,37)(280,37)
\ArrowLine(250,87)(220,37)
\ArrowLine(280,37)(250,87)
\Text(245,107)[lb]{\large{$b_l$}}
\Text(305,12)[lb]{\large{$v_n$}}
\Text(185,12)[lb]{\large{$v_m$}}
\Text(275,62)[lb]{\large{$\chi,\chi'$}}
\Text(242,6)[lb]{\large{$\mbox{(b)}$}}
\Line(420,87)(420,97)
\ArrowLine(450,37)(470,27)
\ArrowLine(370,27)(390,37)
\Line(420,87)(390,37)
\Line(390,37)(450,37)
\Line(450,37)(420,87)
\Text(415,107)[lb]{\large{$M^*$}}
\Text(475,12)[lb]{\large{$\lambda$}}
\Text(355,12)[lb]{\large{$\lambda$}}
\Text(445,62)[lb]{\large{$A'$}}
\Text(414,6)[lb]{\large{$\mbox{(c)}$}}
\end{picture}
\end{center}
\caption{
{\bf(a)} Anomalous superdiagram related to super-Weyl symmetry.
{\bf(b)} The well-known anomaly triangle diagram which is the component diagram
 of (a).
{\bf(c)} Feynman diagram deriving the gaugino mass which is also the
 component of (a). In this diagram only the PV fields encircle. 
}
\label{feyngaugino}
\end{figure}
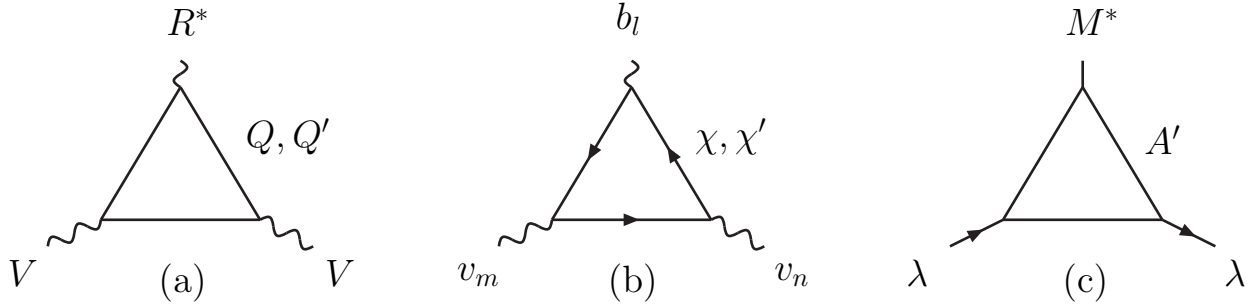

Firstly we reproduce the original SBT's (\ref{AMgauginomass}) and
(\ref{AMscalarmass}) diagrammatically. As usual the super-Weyl anomaly arises
through a triangle diagram. The connection of the super-Weyl
transformation is included in $R$, which is the chiral superfield containing the scalar curvature, and we have necessarily to
consider the $R$-inserted triangle diagram Fig.\ref{feyngaugino}(a). In
the following we use the Pauli-Villars (PV) regularization, and not only
the matter fields $Q_i (=A_i+\sqrt{2}\theta \chi_i +\theta\theta F_i)$, but PV fields $Q'_i$ encircle the triangle in
Fig.\ref{feyngaugino}(a). Cardasso and Ovrut \cite{LopesCardoso:1993sq}
calculated the component diagram Fig.\ref{feyngaugino}(b), in which the
external lines are bosonic component, and derived the super-Weyl anomaly.
If the external lines are gaugino $\lambda$, then only the PV fields
encircle in Fig.\ref{feyngaugino}(c) and would produce gaugino
mass $m_\lambda$ if the auxiliary field $M$ is given v.e.v.

Let us then calculate gaugino mass $m_\lambda$ coming from
Fig.\ref{feyngaugino}(c). The Lagrangian of supergravity coupled with
matter and gauge fields is\footnote{In the following argument we use the notation in Ref.\cite{Wess:cp}.}
\begin{equation}
\mathcal{L} =\mathcal{L}_M+\mathcal{L}_G, \label{wholeL}
\end{equation}
\begin{eqnarray}
\mathcal{L}_M &= &\int \! d^2 \Theta\  2\mathcal{E}
\! \left[ - \frac{1}{8}(\bar{\mathcal{D}}\bar{\mathcal{D}}-8R) \left\{  Q^{\dagger}_i e^{2V} Q_i + Q'^{\dagger}_{i} e^{2V} Q'_{i}\right\} + m'_i\bar{Q}'_iQ'_i \right] + h.c. \label{Lmatter},\\
\mathcal{L}_G &= & \int \! d^2 \Theta\  2\mathcal{E}
\! \left[  \frac{1}{4g^2}  {W^a}^\alpha W^a_\alpha   - \frac{1}{8}(\bar{\mathcal{D}}\bar{\mathcal{D}}-8R)   \Phi'^{\dagger}_{i} e^{2V} \Phi'_{i}  +m'_\Phi \bar{\Phi}'_i \Phi'_i \right] + h.c.
\end{eqnarray}
Here,  $Q'_i$'s  are
chiral PV fields  of the same representation as matter fields
and $\Phi' (= A'_\Phi+\sqrt{2}\theta \chi'_\Phi +\theta\theta F'_\Phi)$
are those of the adjoint representation. 
The PV fields are distinguished by prime " $'$ " from ordinary fields.  $\bar{Q}$ and 
$\bar{\Phi}$ are the conjugate representation chiral fields of $Q$ and $\Phi$
respectively. 
Furthermore, for the sake of simplicity we ignore Yukawa coupling. 

We would like to consider the low energy limit of (\ref{wholeL}). Taking the
flat limit for component fields of gravity multiplet (graviton
$e\rightarrow\delta$, gravitino $\psi\rightarrow 0$, auxiliary fields
$b_n\rightarrow 0,M\rightarrow 0$), eq.(\ref{wholeL}) becomes global SUSY
Lagrangian $\mathcal{L}^{global}
=\mathcal{L}_M^{global}+\mathcal{L}_G^{global}$, where 
\begin{eqnarray}
\mathcal{L}^{global}_M &= &\int \! d^4 \theta\  
\! \left[   Q^{\dagger}_i e^{2V} Q_i + Q'^{\dagger}_{i} e^{2V} Q'_{i} \right] 
+\int \! d^2 \theta\  m'_i \bar{Q}'_i Q'_i  + h.c. ,\\
 \mathcal{L}^{global}_G &= & \int \! d^2 \theta\  
\!   \frac{1}{4g^2}  {W^a}^\alpha W^a_\alpha   + h.c. \nonumber\\
&&+ \int \! d^4 \theta\   \Phi'^{\dagger}_{i} e^{2V} \Phi'_{i}  
+\int \! d^2 \theta\ m'_\Phi \bar{\Phi}'_i \Phi'_i  + h.c.
\end{eqnarray}

However, the auxiliary field $M$ can have v.e.v. if SUSY is broken down by endowing hidden sector fields with v.e.v.. In this case we can take the limit,  $e\rightarrow\delta$, $\psi\rightarrow 0$,$b_n\rightarrow 0,M\nrightarrow 0$, to obtain
\begin{eqnarray}
\mathcal{L} \rightarrow \mathcal{L}^{global} + 
     \left[ -m'_i \frac{M}{3}^* \! \bar{A}'_iA'_i -m'_\Phi \frac{M}{3}^* \! \bar{A}'_\Phi A'_\Phi  + h.c. \right].
\label{vevlimit}
\end{eqnarray}
Here the terms proportional to $M^*$ appear in addition to the global SUSY
Lagrangian $\mathcal{L}^{global}$. The point to which we should 
pay attention is that the effect of $M$ appears only in the PV fields
part. This means that, on the classical level, there is no signature of
SUSY breaking in the sector of the ordinary fields. On the quantum
level, however, the effect of the v.e.v. of $M$ is communicated through
Fig.\ref{feyngaugino}(c) to the mass of gaugino. This is the mediation
of the super-Weyl anomaly, and  Fig.\ref{feyngaugino}(c) gives us 
\begin{eqnarray}
M_\lambda = \frac{g^2}{(4\pi)^2} \left[ \sum_j T_{R_j} -3C_G \right] \frac{M}{3} ^*, \label{lowestgauginomass}
\end{eqnarray}
where $C_G$ is the value of the quadratic Casimir operator of the adjoint representation and $T_R$ is the Dynkin index associated with the matter representation. This is the lowest order term of (\ref{AMgauginomass}).

\begin{figure}[t]
\begin{center}
\begin{picture}(210,80)(0,15)
\SetWidth{1.0}
\Line(100,55)(80,75)
\Line(120,55)(140,75)
\CArc(110,35)(22.36,117,477)
\Photon(90,25)(70,5){2.5}{2}
\Photon(130,25)(150,5){2.5}{2}
\Line(50,5)(170,5)
\Text(70,85)[lb]{\large{$R^*$}}
\Text(140,85)[lb]{\large{$R$}}
\Text(175,0)[lb]{\large{$Q$}}
\Text(25,0)[lb]{\large{$Q^\dag$}}
\Text(145,15)[lb]{\large{$V$}}
\Text(105,15)[lb]{\large{$Q'$}}
\end{picture}
\end{center}
\caption{
Feynman diagram deriving the scalar mass if the external
matter fields are the scalars $A_i$.
}
\label{feynscalar}
\end{figure}
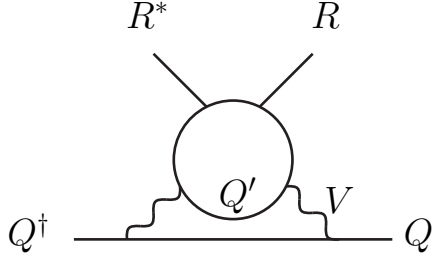

The masses $m_i$'s of the scalar fields $A_i$ are also reproduced similarly. In
this case, however, the scalar masses arise on the two-loop level through
the diagram of the type depicted in Fig.\ref{feynscalar}. If the external
matter fields in Fig.\ref{feynscalar} are the scalars $A_i$, the direct
evaluation gives us the scalar masses,
\begin{eqnarray}
m^2_i =\frac{g^4}{(4\pi)^4} C_R\left[3C_G -\sum_j T_{R_j} \right]\left| \frac{M}{3}\right|^2. \label{lowestscalarmass}
\end{eqnarray}
These results agree in the lowest order with (\ref{AMscalarmass}).

\subsection{K\"{a}hler part}\label{Kahlerpart}
To shed light on the contribution from K\"{a}hler part of the super-Weyl-K{\" a}hler anomaly in the same way as in the above graphical analysis, we  rewrite (\ref{Lmatter}) by using the general K\"{a}hler potential, $K$.
\begin{eqnarray}
\mathcal{L}_M &= & \frac{1}{\kappa^2} \int \! d^2 \Theta\  2\mathcal{E}
\! \left[ \frac{3}{8}(\bar{\mathcal{D}}\bar{\mathcal{D}}-8R) \exp\left\{-\frac{\kappa^2}{3}K \right\} +m\bar{Q}'_i Q'_i \right] + h.c. \\
&&K= Q^{\dagger}_i e^{2V} Q_i + Q'^{\dagger}_{i} e^{2V} Q'_{i}.\nonumber
\end{eqnarray}
Applying the low energy limit while keeping non-vanishing $M$ and taking those on the order of $\kappa^2$, (\ref{vevlimit}) is replaced by
\begin{eqnarray}
\mathcal{L} \rightarrow  \mathcal{L}^{global} + \left[ -m'_i \left\{\frac{M}{3}^* - \frac{2}{3} \kappa^2 K_iF^i \right\}\! \bar{A}'_i A'_i -m'_\Phi \frac{M}{3}^* \! \bar{A}'_\Phi A'_\Phi + h.c. \right], \label{LlowenargylimitWK}
\end{eqnarray}
where, $K_i= \frac{\partial K}{\partial A_i}$. If  hidden sector fields
have v.e.v., the term $\kappa^2 K_iF^i$ in (\ref{LlowenargylimitWK}) containing K\"{a}hler potential has v.e.v. of the same order as $M$. The effect of K\"{a}hler potential also appears in the term containing only PV fields in eq.(\ref{LlowenargylimitWK}). However there is a difference worth remarking, i.e., the term $\kappa^2 K_iF^i$ appears only in the matter PV fields $Q'_i$, but not in the gauge PV fields $\Phi'_i$.  Hence including the effect of the K\"{a}hler symmetry amounts to replacing $\frac{M}{3}^*$ by the following way
\begin{eqnarray}
\frac{M}{3}^* \rightarrow \frac{M}{3}^* - \frac{2}{3} \kappa^2 K_iF^i, \label{replacing}
\end{eqnarray}
without touching upon $\frac{M}{3}^*$ in the $\Phi'_i$ part. Note that the term $T_{R_j}$ in (\ref{lowestgauginomass}) and (\ref{lowestscalarmass}) comes from the matter PV fields $Q'_i$ and  the term $3C_G$ from $\Phi'_i$. We are thus led to formulas
\begin{eqnarray}
M_\lambda &= &\frac{g^2}{(4\pi)^2} \left[  -3C_G \frac{M}{3}^*  + \sum_j T_{R_j} \left\{ \frac{M}{3}^* - \frac{2}{3} \kappa^2 K_iF^i \right\} \right] \label{gauginomass},\\
m^2_i &= & \frac{g^4}{(4\pi)^4} C_R\left[3C_G  \left| \frac{M}{3}\right|^2 -\sum_j T_{R_j} \left| \frac{M}{3}^* - \frac{2}{3} \kappa^2 K_iF^i \right|^2\right], \label{scalarmass}
\end{eqnarray}
which are more general than  (\ref{lowestgauginomass}) and (\ref{lowestscalarmass}) and encompass the super-Weyl-K\"{a}hler anomaly.
The formula (\ref{gauginomass}) agrees with the results obtained previously in Refs.\cite{Bagger:1999rd} and \cite{Gaillard:2000fk}.
The two-loop result (\ref{scalarmass}) for the scalar masses is our new formula.

\section{Tachyonic Slepton Problem}\label{canbesolved}

Let us consider the tachyonic slepton problem on the basis of our new formula (\ref{scalarmass}).
Our following discussions differ for the case of gauge non-singlet scalar from the singlet case, such as right handed slepton in MSSM. This is because the first term in the brackets of (\ref{scalarmass}) is absent for the gauge singlet case. Let us begin with the non-singlet case. On looking at our formulae (\ref{scalarmass}), we immediately notice the possibility that the second negative contribution could  be rendered very small in contrast to (\ref{lowestscalarmass}) by choosing
\begin{equation}
\frac{M}{3}^* - \frac{2}{3} \kappa^2 K_iF^i \approx 0,\label{positivecondition}
\end{equation}
thereby making $m_i^2$ positive. We now argue that eq.(\ref{positivecondition}) is realized in fact quite possibly in general class of models. As a toy model, we take the following simple O'Raifeartaigh model in the hidden sector, whose superpotential is given by
\begin{equation}
P = aX(Z^2-m_{hid}^2) + bm_{hid}YZ.
\end{equation}
Here $a$ and $b$ are complex dimensionless parameters, $m_{hid}$ is a typical mass parameter in the hidden sector and $X,Y$ and $Z$ are chiral superfields. According to \cite{Wess:cp}, $M^*$ is given, by virtue of  equations of motion, in terms of the superpotential,
\begin{eqnarray}
\frac{M^*}{3} &= &-\kappa^2 e^{\kappa^2 K/2}  P^*.  
\label{Mvalue}
\end{eqnarray}
For example, suppose that the mass scale of the hidden sector is on the order of $10^{14}$ GeV. It then turns out that $\kappa^2 K$ is negligibly small and we are able to express (\ref{Mvalue}) as a polynomial of $a$ and $b$.  If these parameters satisfy $b \approx \frac{\sqrt{20}}{3}ia$ ($a=$ real) then we find eq.(\ref{positivecondition}) is realized. We have thus found a possibility to get around the tachyonic slepton problem. 

Until now we have assumed that the contribution of anomaly mediation is more dominant than that of gravity mediation.
One may perhaps wonder whether we also have to include the effect due to gravity mediation to the slepton mass in so far as the hidden and visible sectors are bridged. We are, however, still able to choose the  K\"{a}hler potential and superpotential in such a way that the gravity mediation effects can be neglected.

Let us turn to the singlet case. If we demand (\ref{positivecondition}), the scalar mass
squared (\ref{scalarmass}) is zero because $C_G$ term is absent, and
this situation is unrealistic. To overcome the drawback we extend our
model a little further. Original anomaly mediation has the property to
be insensitive to ultraviolet physics. The induced SBT's
are independent of fields which have heavy masses. The heavy mass
threshold effect dose not appear. 
Extending the model, however, we can
evade this property. Nelson and Weiner \cite{Nelson:2002sa} have introduced a bilinear
coupling in the K{\" a}hler potential and have seen
violation of the insensitivity due to the threshold correction. We now take
their idea into our consideration of the singlet case. Due to bilinear
terms the ultraviolet insensitivity is spoiled, so our prediction can be
possibly modified. (Adding bilinear couplings, a mass scale could be introduced, whose mass scale looks like the Higgs $\mu$ parameter. We may get, however, an undesirable prediction of the so-called $B$ term, and we leave it open for now.)

To have an insight into the effect of bilinear coupling, let us recall
the essence of the ultraviolet insensitivity. The low energy limit of a field $W$ with the heavy mass $m_{W}$ is  similar to  (\ref{LlowenargylimitWK}), 
\begin{eqnarray}
\mathcal{L}_W \rightarrow  \mathcal{L}^{global}_W + \left[ -m_{W} \left\{\frac{M}{3}^* - \frac{2}{3} \kappa^2 K_iF^i\right\}\! \bar{A}_W A_W + h.c. \right]. \label{Llowenargylimitheavy}
\end{eqnarray}
If $m_{W}$ is large enough, the effect from heavy fields is the same as
the one from PV fields and hence is equivalent to adding another PV
field. As far as the effect is independent of the regularization, the
heavy field effect is not seen. 
In this sense even if the K{\" a}hler potential term is included, the insensitivity is preserved.
Thus heavy fields effect vanish and anomaly mediation is insensitive to physics in the ultraviolet.

The insensitivity is , however, no longer true, if we add two chiral
fields $H$ and $\bar{H}$ with bilinear coupling. The part of the K{\" a}hler potential of additional fields is 
\begin{equation}
K_H = H^\dagger H   +  \bar{H}^\dagger \bar{H} - c \bar{H}H- c^* \bar{H}^\dagger H^\dagger,\label{addtionalK}
\end{equation}
where $c$ is the parameter describing the bilinear coupling. Applying
the same limit as in (\ref{vevlimit}), the Lagrangian of additional fields is
\begin{eqnarray}
\mathcal{L}_H \rightarrow \mathcal{L}^{global}_H + \left[ -m_{c} \left\{-\frac{M}{3}^* - \frac{2}{3} \kappa^2 K_iF^i \right\}\! \bar{A}_h A_h  + h.c. \right] ,\label{Llowenargylimitthreshold}
\end{eqnarray}
where $m_{c}= \frac{c M}{3}$.
On looking at (\ref{Llowenargylimitheavy}) and
(\ref{Llowenargylimitthreshold}), we immediately notice a remarkable
difference in sign in front of  $\frac{M}{3}^*$.  Because of this
difference, the heavy fields $H$ and $\bar{H}$ discriminate themselves
from PV fields and the threshold effects survive the low energy limit.
At the threshold $m_c$, below which the additional fields decouple, the
$U(1)$ gaugino mass $M_{U(1)}$ for example is given by
\begin{equation}
M_{U(1)} = \frac{\alpha(m_{c})}{4\pi} \left[ 2 \sum \left( T_{R_H}+T_{R_{\bar{H}} } \right) \frac{M}{3}^* \right] .
\label{gauginoatmc}
\end{equation}
Here we have included an arbitrary number of $(H,\bar{H})$ pairs and the summation in (\ref{gauginoatmc}) is taken over the pairs.
where the sum is over additional fields.
Because of the threshold correction, SBT's are no longer on the anomaly
mediation trajectory. We must take account of the running effect from
$m_{c}$ to $\mu$. Solving  renormalization group equation, we find that
the $U(1)$ gaugino mass and the singlet scalar mass $m^2_{singlet}$ at
the energy scale $\mu$ are  
\begin{eqnarray}
M_{U(1)} &=& \frac{\alpha(\mu)}{4\pi} \left[ 2 \sum \left( T_{R_H}+T_{R_{\bar{H}} } \right) \frac{M}{3}^* \right] \label{gauginomassthreshold},\\
m^2_{singlet} &=& \frac{8 C_R}{(4\pi)^2} \left[\frac{ \left\{ \sum \left( T_{R_H}+T_{R_{\bar{H}} } \right)\right\}^2 }{ \sum_j T_{R_j}}
           \left\{\alpha^2(m_{c}) -  \alpha^2(\mu) \right\} \right.  \nonumber\\ && \hs{35} \left.
          -\frac{1}{2}\sum \left( T_{R_H}+T_{R_{\bar{H}} } \right)\alpha^2(m_{c})
		   \right]  \left| \frac{M}{3}\right|^2 .\label{scalarmassthreshold}
\end{eqnarray}
These formulae replace the previous ones, (\ref{gauginomass}) and (\ref{scalarmass}).

In order to see if the threshold corrections are large enough, let us look for conditions to be imposed on the number of $(H,\bar{H})$ pairs. 
We assume $\frac{M}{3} \sim $100TeV which guarantees that the gaugino masses are on the order of 1TeV.
We also assume $c\sim \mathcal{O}(1)$ and $\mu \sim $100GeV.
We then find $\sum \left( T_{R_H}+T_{R_{\bar{H}} } \right)>$22 is the necessary condition to ensure $m_{singlet} >100$GeV for the case that $H$ and $\bar{H}$ are all SU(2) singlet.
In other words, if the number of $(H,\bar{H})$ pairs is greater than eleven, we get positive scalar masses.
Finally we remark that the masses of the gauge non-singlet scalars are not much affected by the threshold effects.
We can in fact easily confirm that the contributions (\ref{scalarmass}) to the non-singlet scalars are more dominant numerically for $\sum \left( T_{R_H}+T_{R_{\bar{H}} } \right)= 24$ than the threshold effects.
 
 \section{Conclusion}
 
Let us summarize what we have done in this letter. The theory of supergravity coupled with chiral matters and gauge multiplet is invariant under the super-Weyl-K{\" a}hler transformation, and the  symmetry  has the anomaly to be broken down on the quantum level. Our diagrammatical analysis has shown that the K{\" a}hler part of the anomaly is mediated in parallel with the super-Weyl part which has been studied previously. The SBT's which have been known to contain $M$ in the anomaly mediation scenario, have now new contribution driven by K{\" a}hler part of anomaly and the whole effect is summarized by (\ref{gauginomass}) and  (\ref{scalarmass}).

If the condition (\ref{positivecondition}) is satisfied, the mass  squared of the non-singlet scalar such as left handed slepton can be positive. In the singlet scalar case, however, the formulae (\ref{scalarmass}) combined  with (\ref{positivecondition}) leads to a vanishing mass squared $m_{singlet}^2\approx 0$, which is of course unfavorable.  To circumvent such a prediction, we have taken account of the threshold correction, by which deviation from the anomaly mediation trajectory is introduced. We have seen that the threshold corrections elevate the masses of gaugino and gauge singlet scalar, for example $U(1)$ case, as in (\ref{gauginomassthreshold}) and  (\ref{scalarmassthreshold}).
We have thus got the possibility of solving the tachyonic slepton problem.


\vs{10}
\noindent
{\Large {\bf Acknowledgement}}
\\ \\
\noindent
We would like to thank Prof. B.~Nelson for calling our attention to Refs.\cite{Gaillard:2000fk} and \cite{Binetruy:2000zx}.
\noindent
The work of T.K. is supported in part by Grant in Aid 
for Scientific Research from the Ministry of Education 
(grant number 14540265 and 13135215). 


\end{document}